\documentclass{article}
\usepackage{spconf,amsmath,graphicx}

\usepackage{enumitem} 
\setlist{nosep, leftmargin=14pt}

\usepackage{mwe} % to get dummy images

\usepackage{color,soul}
\usepackage{algorithm}
\usepackage{algorithmic}
\usepackage{mathrsfs}
\usepackage{array}
\usepackage{mdwmath}
\usepackage{mdwtab}
\usepackage{eqparbox}
\usepackage{url}
\usepackage{float}
\usepackage{amsmath}
\usepackage{graphicx}
\usepackage{sidecap}
\usepackage{enumitem} % to remove the indent of itemize/enumerate
 
\usepackage{blindtext}

\usepackage{subfiles} % Best loaded last in the preamble

\usepackage{todonotes}

\title{YOLO2U-Net: Detection-Guided 3D Instance Segmentation for Microscopy}
\twoauthors
 {Amirkoushyar Ziabari\thanks{This manuscript has been authored by UT-Battelle, LLC, under Contract No. DEAC05-00OR22725 with the U.S. Department of Energy. The United States Government and the publisher, by accepting the article for publication, acknowledges that the United States Government retains a non-exclusive, paid-up, irrevocable, world-wide license to publish or reproduce the published form of this manuscript, or allow others to do so, for United States Government purposes. DOE will provide public access to these results of federally sponsored research in accordance with the DOE Public Access Plan (http://energy.gov/downloads/doe-public-access-plan). 
 Corresponding author's email address: ziabariak@ornl.gov}, Derek C. Rose}
 {Mutlimodal Sensor Analytics Group\\
	Oak Ridge National Lab (ORNL)\\
	Oak Ridge, TN 37830}
  {Abbas Shirinifard, David Solecki}
	{St. Jude Children's Hospital \\
	Memphis, TN, 37923 }
\begin{document}
%\ninept
%
\maketitle
\begin{abstract}
Microscopy imaging techniques are instrumental for characterization and analysis of biological structures. 
As these techniques typically render 3D visualization of cells by stacking 2D projections, issues such as out-of-plane excitation and low resolution in the $z$-axis may pose challenges (even for human experts) to detect individual cells in 3D volumes as these non-overlapping cells may appear as overlapping. 
In this work, we introduce a comprehensive method for accurate 3D instance segmentation of cells in the brain tissue. 
The proposed method combines the 2D YOLO detection method with a multi-view fusion algorithm to construct a 3D localization of the cells. 
Next, the 3D bounding boxes along with the data volume are input to a 3D U-Net network that is designed to segment the primary cell in each 3D bounding box, and in turn, to carry out instance segmentation of cells in the entire volume. 
The promising performance of the proposed method is shown in comparison with current deep learning-based 3D instance segmentation methods.
\end{abstract}

\begin{figure*}
\begin{center}
\includegraphics [scale=.28,trim=.5cm 0cm 0cm 0cm,clip]{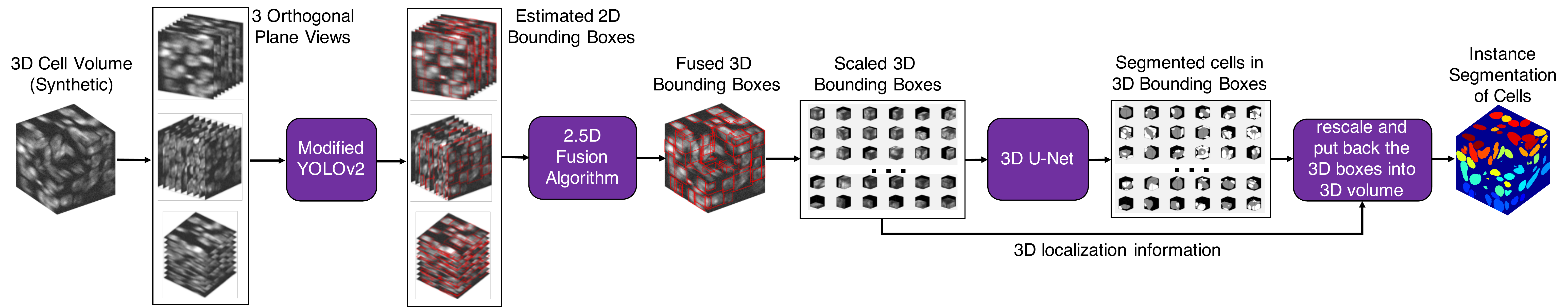}
\end{center}
\caption{YOLO2U-Net architecture. }
\label{YOLO2UNet}
\end{figure*}

\vspace{-.25cm}
\section{Introduction}
\label{sec:intro}

Advances in training deep convolutional neural networks have driven development in network architectures that are focused on semantic segmentation to pixel-wise label images. %(COCO~\cite{MS_COCO2014} and PASCAL VOC~\cite{PASCAL2014})
These approaches have largely focused on natural image segmentation (COCO~\cite{MS_COCO2014} and PASCAL VOC~\cite{PASCAL2014}), though encoder-decoder networks with skip connections \cite{ronneberger_unet_2015} and other multi-scale techniques (spatial pyramids \cite{zhao_pyramid_2017} and atrous convolution or pooling \cite{chen_rethinking_2017}) have also shown success across a variety of medical and other imagery.
%These approaches have largely focused on natural image segmentation (COCO and PASCAL VOC), though encoder-decoder networks with skip connections \cite{ronneberger_unet_2015} and other multi-scale techniques (spatial pyramids \cite{zhao_pyramid_2017} have also shown success across a variety of medical and other imagery.

Biomedical image analysis presents challenges which are somewhat unique in instance segmentation.  
The orientation and concentration of objects can be random, objects can appear at varying scales, object boundaries can be unclear or overlapping, and object texture can vary spatially and contextually. 
Low contrast and noise or imaging artifacts such as out-of-plane excitation can make separating objects tedious and difficult to automate, often leading to a shortage of labeled data. 
Biomedical data sets are often inherently 3D, though potentially highly anisotropic with lower depth-wise resolution. 
Our work is motivated by the segmentation of nucleui in 3D microscopy volumes. 
While Fully Convolutional Neural (FCNs)~\cite{long_fully_2015} have been extensively used to develop state-of-the-art 2D instance segmentation algorithms~\cite{ronneberger_unet_2015,chen_rethinking_2017, He_2017_ICCV, Vuola8759574}, they are mainly suitable for 2D instance segmentation of natural scenes; and/or are computationally expensive for 3D segmentation of a full image volume.  We emphasize that, in this work, 3D explicitly implies the \textit{dimensionality} of an object in an image volume and not the \textit{depth} of an object in a 2D image.

A number of approaches have been taken to address computational limitations in 3D segmentation, spanning from integrating tri-planar views to recurrent neural networks for capturing slice to slice context \cite{chen_combining_2016}. 
3D convolution with a U-Net topology for relatively small volumes (order of 100x100x100 voxels) was performed successfully on biomedical imagery by \cite{dou_3d_2016} and \cite{cicek_3d_2016}.  
Recently, ~\cite{ dunn2019deepsynth} developed DeepSynth, which combines SpCycleGAN, used to generate synthetic cell data for training, with a modified 3D U-Net network to 3D segment real cell data. 
DeepSynth encompasses a slice-by-slice based watershed and morphological post-processing algorithm for instance segmentation of touching cells.
DeepCell ~\cite{bai2017deep, wang2019learn} is another state-of-the-art method for instance image segmentation of volumes containing overlapping cells. 
DeepCell's deep learning-based watershed segmentation approach can handle overlapping cells in noisy volumes without over-segmentation, which is a typical drawback of traditional watershed algorithms. 
Another recent approach is StarDist that allowes 3D segmentation of cells limited to star-convex shaped objects~\cite{weigert2020star}.

In this paper, we propose a comprehensive detection and segmentation framework, called YOLO2U-Net, that judiciously combines two successful network topologies, namely You-Only-Look-Once (YOLO) \cite{redmon2017yolo9000} and U-Net \cite{cicek_3d_2016}, to do 3D object detection and instance segmentation for cells in microscopy volumes.
In the proposed method, we first modify and use YOLO~\cite{redmon2017yolo9000} to detect cells from 2D orthogonal perspectives of a 3D volume and plot a bounding box around each one of them. 
We then follow with an algorithm that combines the 2D detected bounding boxes to localize the cells in 3D bounding boxes.
Finally, instance-level segmentation of 3D cells within the detected 3D bounding boxes is performed using a 3D U-Net~\cite{cicek_3d_2016} modified for unbalanced data. 
This 3D network segments out the primary cells in each bounding cube. 
The proposed method is an extension to our work in ~\cite{ziabari2019TwoTierCNN}.
In comparison the proposed method removes the necessity of any post-processing watershed and morphological operations for separating the cells; rather the entire cell  will be segmented out in each 3D bounding box.
This in turn avoids common artifacts when performing instance segmentation associated with stitching sub-volumes together such as over-segmentation of cells and missing cell-cell boundaries (especially for high cell confluence cases).

In section~\ref{method}, we describe the proposed method and its comprising components. Section~\ref{results} include synthetically generated data along with metrics, experimental and quantitative comparisons with the state-of-the-art methods. We conclude the paper and propose future plans in section~\ref{conclusions}.

\begin{algorithm}
\vspace{-.05cm}
\caption{2.5D YOLO-Based Fusion Algorithm\label{alg1}}
\begin{algorithmic}
\scriptsize
%\STATE Perform YOLOv2 detection on sagittal, coronal, and xial slices
%\STATE \qquad 
\STATE 1: Predict all bounding boxes for 2D cells in X-Y, X-Z, Y-Z views (YOLOv2~\cite{redmon2017yolo9000})
\STATE 2: Reject bounding boxes with less than $50\%$ confidence 
\STATE 3: Pair-wise comparison between bounding boxes from different views 
\STATE 4: Estimate the coordinates of proposal 3D bounding boxes \\ 
        $[x_{min,est}, y_{min,est}, z_{min,est}, x_{max,est}, y_{max,est}, z_{max,est}]$
        %For X-Y view (arbitrary starting point):
        %- For each box get the $[x_{min}, x_{max}, y_{min}, y_{max}, z]$   
		%- Find cluster of boxes which intersect among other X-Z, Y-Z views
\STATE 5: Cluster 3D bounding boxes with more than $5\%$ overlap.
\STATE 6: Estimate $[x_{min,est}, y_{min,est}, z_{min,est}, x_{max,est}, y_{max,est}, z_{max,est}]$ \\ 
          (We use the median of the merged cluster of boxes, i.e. based on median value of coordinates in each set we obtain a single 3D bounding box per cluster). 
\STATE 7: Non-Maximum Suppression to avoid duplicate 3D bounding boxes
			%- How do we come up with a single confidence score given each view/perspective (and each box) has its own confidence? Options: median, mean, ?
			%- Without this confidence score, we only can calculate precision and recall for a given IoU with ground truth, no sorting for mAP calculation

\end{algorithmic}
%\captionsetup{justification =  centering, margin =2cm}
\end{algorithm}
\vspace{-0.65cm}
\raggedbottom

\section{YOLO2U-Net}\label{method}

%In this section we explain the methods used in this work to analyze the data and to perform the comparisons. First, we explain the proposed method and provide details on its main building blocks. 
%We explain two variety of the proposed method, Two-Tier CNN and YOLO2U-Net, stemming from different training strategy used in the U-Net step of the proposed method. Further, we introduce the DeepCell~\cite{bai2017deep, wang2019learn} as the reference state-of-the-art method to compare against the proposed approaches.

%\subsection{YOLO2U-Net}

Figure~\ref{YOLO2UNet} summarizes the proposed method.
In this method, we first localize the cells using a 2.5D fusion algorithm that fuses the 2D bounding boxes of the cells that are obtained from 2D orthogonal perspectives of the 3D volume of cells.
The YOLO~\cite{redmon2017yolo9000} network is used to obtain 2D bounding boxes in each perspective.
% (additional details on YOLO are provided in the SI).
The volume of data with localization information are then input to 3D U-Net~\cite{cicek_3d_2016}, which is trained to identify the main cells in each 3D bounding box and separate them from portions of the neighboring cells within the same box.
Finally, we leverage the knowledge about the position of individual boxes to put the cells back to the original volume. 
In the following, we describe each component of our proposed method separately. 

\begin{figure}%[t!]
\centering
\includegraphics [scale=.3]{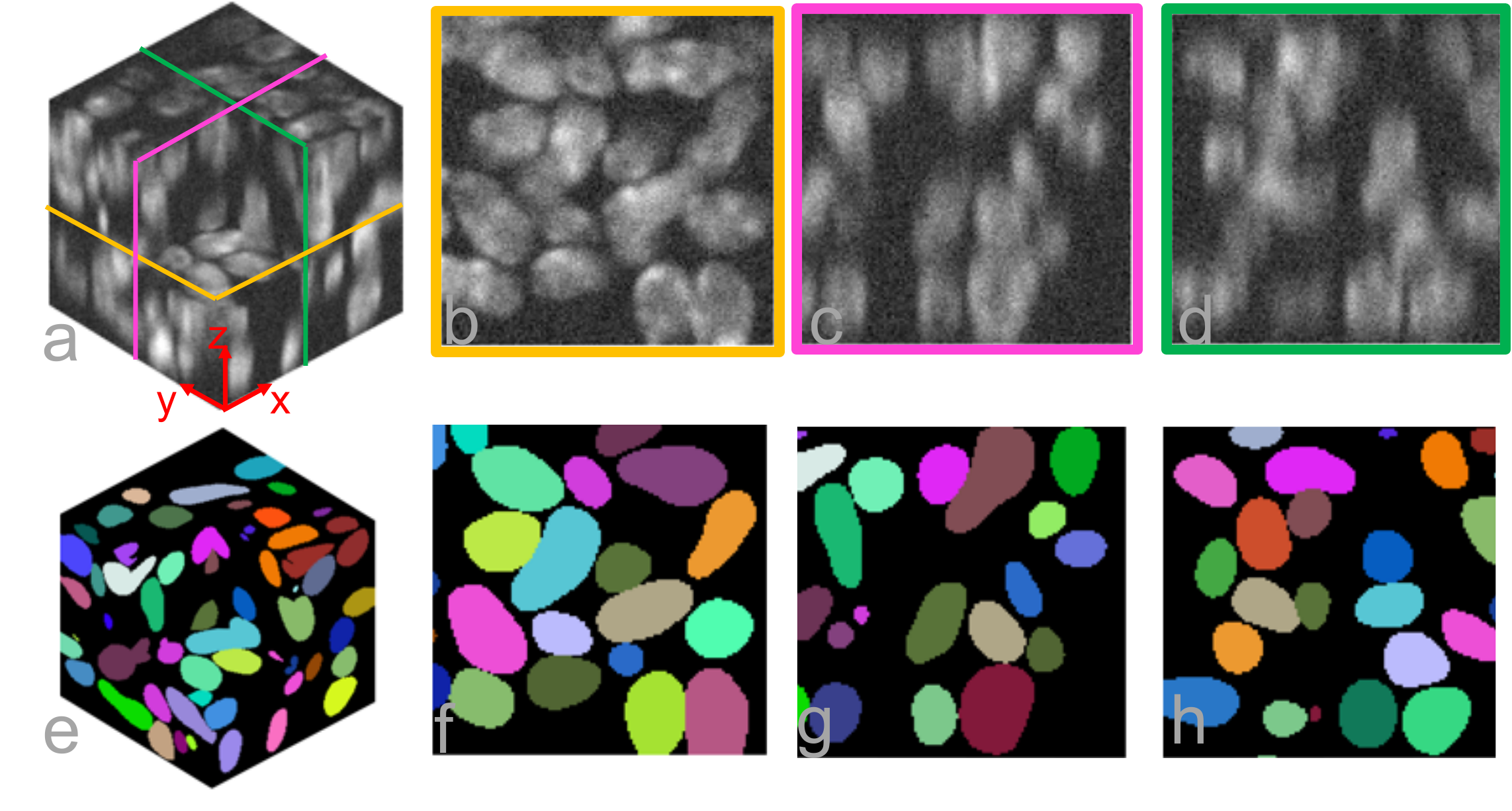}
\caption{An example volume from test data set is shown along with the ground truth instance segmentation.
Three central cross sections along axial ($X-Y$), coronal ($X-Z$), and sagittal ($X-Z$) planes are also shown.}
\label{Inst_Seg_Vols}
\vspace{-0.5cm}
\end{figure}
\raggedbottom

 \vspace{-.25cm}
\subsection{2.5D YOLO-based Fusion Algorithm}

YOLO, and its more recent version YOLOv2~\cite{redmon2017yolo9000}, is a state-of-the-art fast 2D object detection algorithm. 
Despite its utility for fast localization in 2D, it becomes exponentially more expensive to perform 3D localization with 3D convolutions within the YOLOv2 topology. 
To leverage the performance of YOLOv2 for 3D localization of objects in 3D volumes of data we propose the 2.5D fusion algorithm outlined in Algorithm~\ref{alg1}.

In this algorithm, we first train the YOLOv2 network on 2D slices from orthogonal Cartesian planes of a 3D synthetic image volume.
%When the objects are isotropic, training on slices from one of the 2D planes
YOLOv2 may produce several bounding boxes per object; only bounding boxes with more than $50\%$ confidence are kept. 
This step is then followed by a non-maximum suppression to discard boxes that have more than $50\%$ overlap to make sure each object is localized only once. 
For each object that extends in 3D, there must be a nonempty region of intersection between bounding boxes in each view. 
Therefore, all the detected 2D bounding boxes are pairwise compared with other boxes in the same plane as well as with boxes from orthogonal planes. %Our analysis shows that for confidence levels less than $\sim$ $85\%$, the network performs consistently and this parameter has negligible impact on predicted boxes~\cite{ziabari2019AsilLocal}.
The coordinates of the overlapped 2D boxes are joined to obtain the proposal coordinates for the 3D bounding boxes ($[x_{min}, x_{max}, y_{min}, y_{max}, z_{min}, z_{max}]$).
Next, to prune the multiple boxes created from multi-perspective detection, all the proposal 3D bounding boxes are pair-wise compared and those with more than $5\%$ overlap are clustered together.
This threshold was empirically chosen by testing the fusion algorithm on synthetic volumes of touching spheres. % so that 3D bounding boxes within the margin of errors get merged.
%To set this, we investigated the choice of overlap threshold for a set of synthetic data sets (please see SI) numbers between $0$ to $15\%$ and found out that $5\%$ produces the best performance across 20 volumes of train data sets.
%Reducing this threshold results in missing 3D bounding boxes (merging too many 3D bounding boxes) while increasing this number will result in more false alarms (breaking too many 3D bounding boxes). 
%The same threshold was then used for the test data sets.
Finally, to obtain the final 3D bounding boxes' coordinates, for each cluster the median of each of the 6 coordinates of all the extracted 3D bounding boxes are calculated and a non-maximal suppression is applied.

\subsection{3D U-Net for 3D Cell Segmentation inside 3D Bounding Boxes}

Since the cells are localized, their positional information can be used to perform guided bounding box selection to perform 3D instance segmentation with a 3D U-Net.
%The structure of the 3D U-Net architecture used in this work~\cite{cicek_3d_2016} is provided in \todo{formal reference?} supplementary information (SI).
%The goal of the 3D U-Net is to segment out cells \AKZ{as a whole} inside the 3D bounding boxes.
%We adapt two strategies for U-Net to segment cells in bounding boxes which we discuss below.
Here, we train the U-Net to separate the main cell from the remaining voxels in the 3D bounding box. 
%An example is shown in Figure~\ref{TrainStrategy_Comp}c. %instead of performing boundary segmentation, 
To mitigate for cells of differing shape and size, during training we scale 3D input cubes of cells to a fixed size for our 3D U-Net ($48\times48\times48$).
This size is chosen based on the cell sizes encountered in the training data and for faster training of the 3D U-Net.
For example, if a bounding box is of size of $59\times30\times47$, we perform zero-padding to make it $59\times59\times59$ and then scale it to $48\times48\times48$.
Once the segmentation inside the bounding box is performed it is resized back to its original size.
Once the segmentation inside the bounding box is performed it is resized back to its original size.
The proposed one-step instance segmentation strategy avoids the post-processing watershed and morphological filtering such as those in~\cite{dunn2019deepsynth, ziabari2019TwoTierCNN}.
After all the bounding boxes are segmented, they are placed in separate volumes of the size of the original input data at their position. 
We then apply an $argmax$ to this 4D volume to label cells for the final 3D volume. %~\hl{ref. of argmax idea}.
\vspace{-0.25cm}

%The is an extension to the work done in~\cite{ziabari2019TwoTierCNN}, where authors used boundary segmentation\hl{bluh}. Details of that method are explained in the supplementary materials and results obtained by that methods are compared against YOLO2U-Net in the next section. 
%Therefore, a second strategy was developed to address these challenges. 

\section{Experimental Results}\label{results}

\subsection{Data sets}\label{data sets_Synth}

CompuCell3D is an open-source toolkit that is widely used to simulate biological cells and tissues~\cite{swat2012multi} using agent-based methods. 
We used a three-compartment virtual cell to simulate cell nucleus shapes and internal distribution of DNA material. 
These virtual cells are flexible and can easily mimic realistic cell nucleus shape and cell-cell boundaries. 
Euchromatin domains are modeled as two equally sized compartments occupying about 85$\%$ of the total volume of a virtual cell. 
Heterochromatin domains are modeled as multiple compartments (5 to 9) occupying about 15$\%$ of the total volume.  
We initialized 128 cells randomly located in a $84\times84\times84$ lattice. 
We crop the lattice to $64\times64\times64$ to avoid artifacts due to lattice boundary conditions. 
The extent of cell-cell contact and organization of the internal compartments can be adjusted by setting appropriate contact energy parameters. 
To increase cell-cell contact, a negative surface tension is required (and vice versa). 
We transform the 3-compartment virtual cells in the $64\times64\times64$ lattice volume to a realistic synthetic image by: 1) up-scaling and creating smooth cell boundary masks and 2) assigning signal intensity to compartments and applying smooth boundary masks. 
To mimic real microscopy data sets, we applied realistic microscopy aberrations (experimental point spread function approximated by a 3D Gaussian Blur) and added Gaussian noise to the final simulated data.

\begin{figure}[t!]
\centering
\includegraphics [scale=.25,trim=0.68cm 0cm 0cm 0cm,clip]{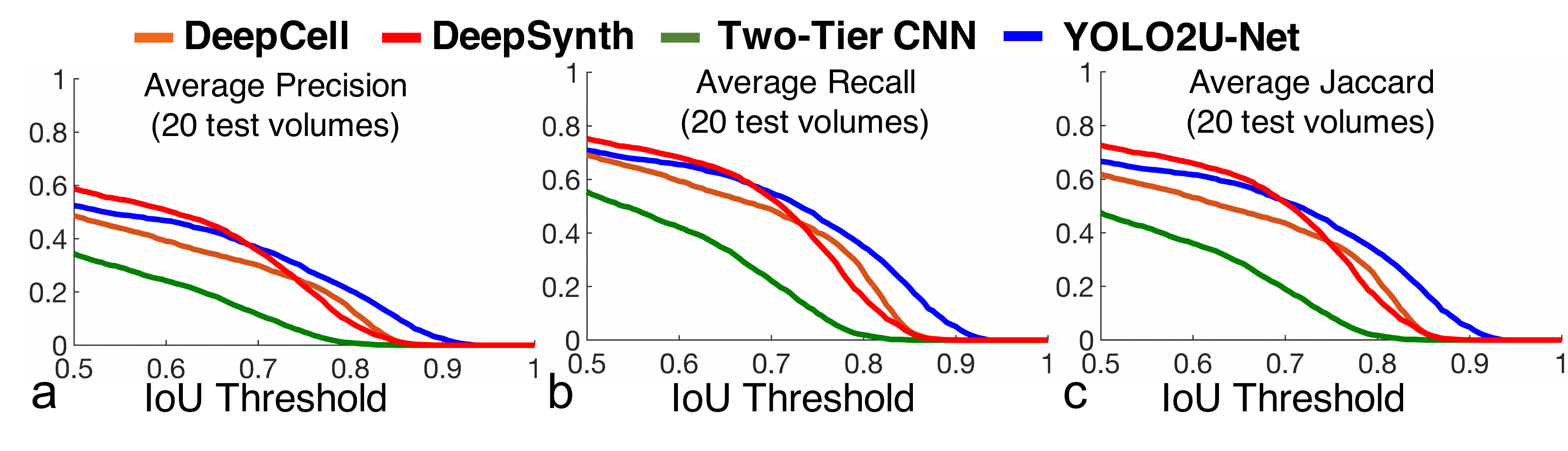}
\caption{Quantitative comparison: ($AP$) ,($AR$), ($AJ$) scores (defined in the text) for 20 volumes against IoU are plotted.}
\label{comp_qunt}
\vspace{-.3cm}
\end{figure}
\raggedbottom

\begin{table}[!htp]
\vspace{-.5cm}
\tiny
%\scalefont{0.5}
\centering
\caption{metric comparison}
\label{Tab1}
\resizebox{\columnwidth}{!}
%\resizebox{.5\textwidth}{!}
{%
\begin{tabular}{|c|c|c|c|c|}
\hline
 & DeepCell~\cite{bai2017deep} & DeepSynth~\cite{dunn2019deepsynth} &  Two-Tier CNN~\cite{ziabari2019TwoTierCNN} &  \textbf{YOLO2U-Net}\\
\hline
mAP & 0.16   & 0.338   & 0.296  & \textbf{0.367}    \\
\hline
mAR & 0.187   & 0.349  & 0.331 & \textbf{0.39}   \\
\hline
mAJ & 0.107   & 0.248 & 0.211  & \textbf{0.263}   \\
\hline
\end{tabular}
}
\vspace{-.7cm}
\end{table}
\raggedbottom

\begin{figure*}[t!]
\centering
\includegraphics [scale=.4]{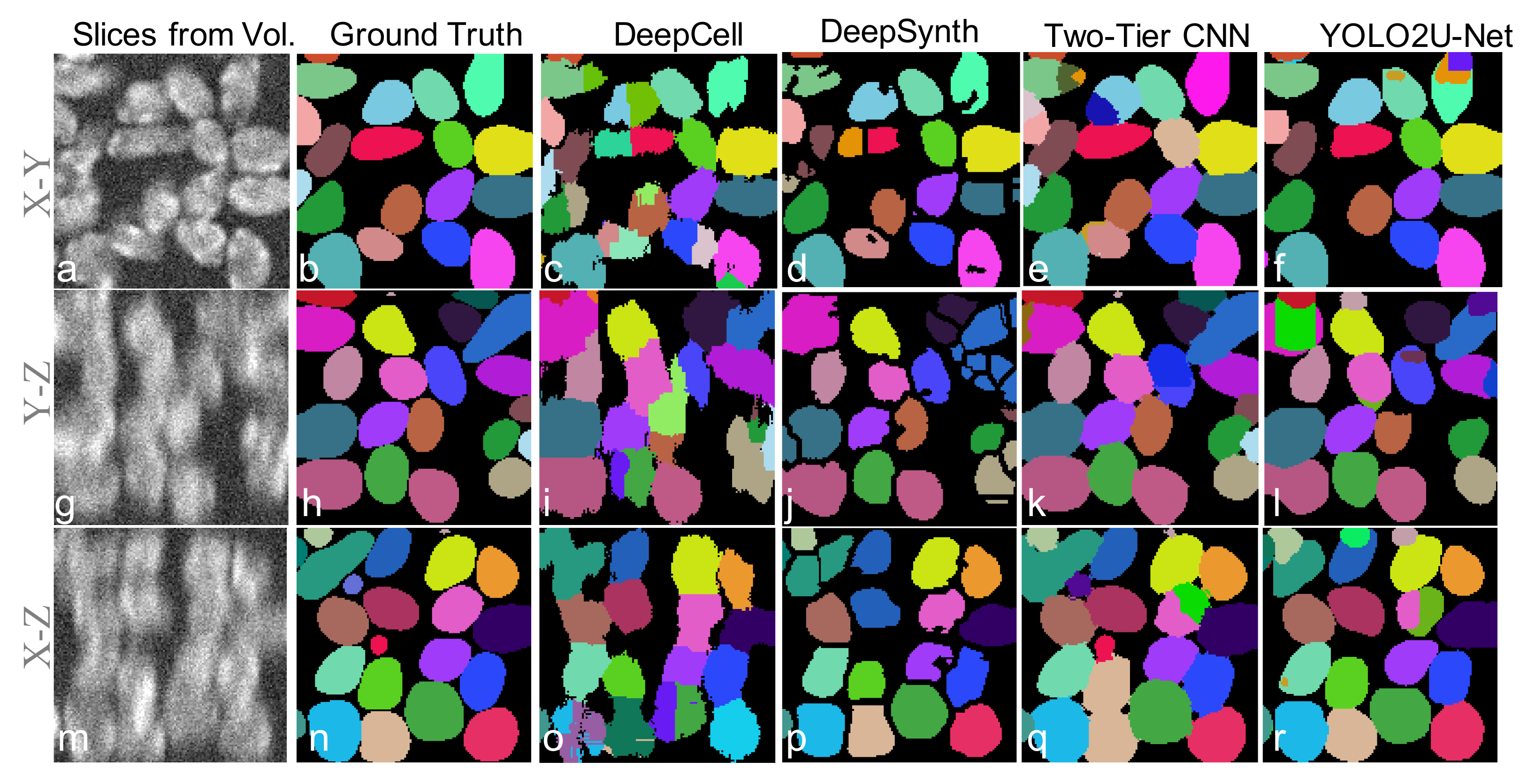}
\caption{Instance segmentation comparison across slices from different views for a volume from test data sets.
Challenging slices are selected to highlight strengths and limitations of the YOLO2U-Net.}
\label{slices_set3}
\vspace{-0.5cm}
\end{figure*}

\subsection{Metrics}

In this section, we summarize the metrics that are used throughout the main text.
The intersection-over-union ($IoU$) between two 3D segmented cells is defined as $ IoU = \frac{Cell_{target} \cap Cell_{predicted} }{Cell_{target} \cup Cell_{predicted} } $, where, $Cell_{target}$ and $Cell_{predicted}$ correspond to voxels of the cell target cell and predicted cell in the 3D volume.
To evaluate the instance segmentation performance, we calculate precision ($P$), recall ($R$) and Jaccard ($J$) scores at the voxel level and as a function of 3D IOU threshold values ($th$) using:
\vspace{-0.3cm}
\begin{align}
\begin{split}
P(th) = \frac{N_{TP}(th)}{N_{TP}(th)+N_{FP}(th)}, \\
R(th) = \frac{N_{TP}(th)}{N_{TP}(th)+N_{FN}(th)}, \\ 
J(th) =  \frac{N_{TP}(th)}{N_{TP}(th)+N_{FP}(th)+N_{FN}(th)}.
\end{split}
\end{align}
Here, $N$ is the number of voxels and $TP$, $FP$, and $FN$ are true positive count, true negative count, and false negative count. 
The average values for $N_{vols}$ test volumes are calculated at each 3D IOU threshold level as average precision $AP$, average recall $AR$ and average Jaccard $AJ$ scores.
By integrating these values over the entire range of IOU levels, we get their mean average precision ($mAP$), mean average recall ($mAR$) and mean average Jaccard ($mAJ$) score.

\subsection{Comparison with State-of-the-art Methods}

We compare YOLO2U-Net against three state-of-the-art methods, namely, DeepCell~\cite{bai2017deep}, DeepSynth~\cite{dunn2019deepsynth}, and our previous work two-tier CNN~\cite{ziabari2019TwoTierCNN}.
To perform a fair comparison, all methods were trained on the same training data sets, which is 20 volumes of $128^3$ simulated cell microscopy data as detailed in section~\ref{data sets_Synth}.

The trained network, then tested on 20 new volumes of cell data that are the same size as training volumes but of course not seen in the training. 
An example of test volumes along with the ground truth instance segmentation mask are shown in Figure~\ref{Inst_Seg_Vols}.
Figure~\ref{comp_qunt}, compares precision, recall, and the Jaccard scores obtained by different methods for test data sets.
Each panel in this Figure plots an average score for 20 volumes in the test data set as a function of 3D intersection-over-union (IoU) levels (in range $\in [0.5,1]$).
We note that YOLO2U-Net performs best among the tested methods in all cases for IOU larger than 0.7 and has about the same performance as DeepSynth at lower IOUs.
The better performance of our method at higher IOUs is indicative of the importance of 3D localization performed prior to instance segmentation.

To further demonstrate the impact of 3D localization, we have shown slices of segmented cells in a test volume from different views in Figure~\ref{slices_set3}.
For these examples, and in particular in X-Z and Y-Z slices, where blurring and out-of-plane excitation worsens the image quality, DeepCell fails to accurately segment boundaries, DeepSynth suffers from  over-segmentation due to post-processing watershed, two-tier CNN misses some cells, while YOLO2U-Net correctly separates cells even when the boundaries are vague and very blurred.
These observation, again, signifies the importance of the localization step.
We should note that, the slices in Figure~\ref{slices_set3} intentionally selected to contain notable cell configurations that highlight both the strengths and limitations of the YOLO2U-Net approach.

In Table~\ref{Tab1}, we compare the three methods in terms of their mean average precision ($mAP$), mean average recall ($mAR$) and mean average Jaccard ($mAJ$) scores. %~\cite{WinNT}
The largest values are shown in bold. 
In all cases, YOLO2U-Net outperforms other tested methods.
%\vspace{-0.5cm}

\subsection{Ablation Study}

In this section we conduct an ablation study to investigate the impact and contribution of each of the components of the proposed architecture (2D localization, detection box fusion for 3D bounding boxes, and 3D segmentation) on the performance of YOLO2U-Net.

\begin{figure}[t!]
\centering
\includegraphics [scale=.32]{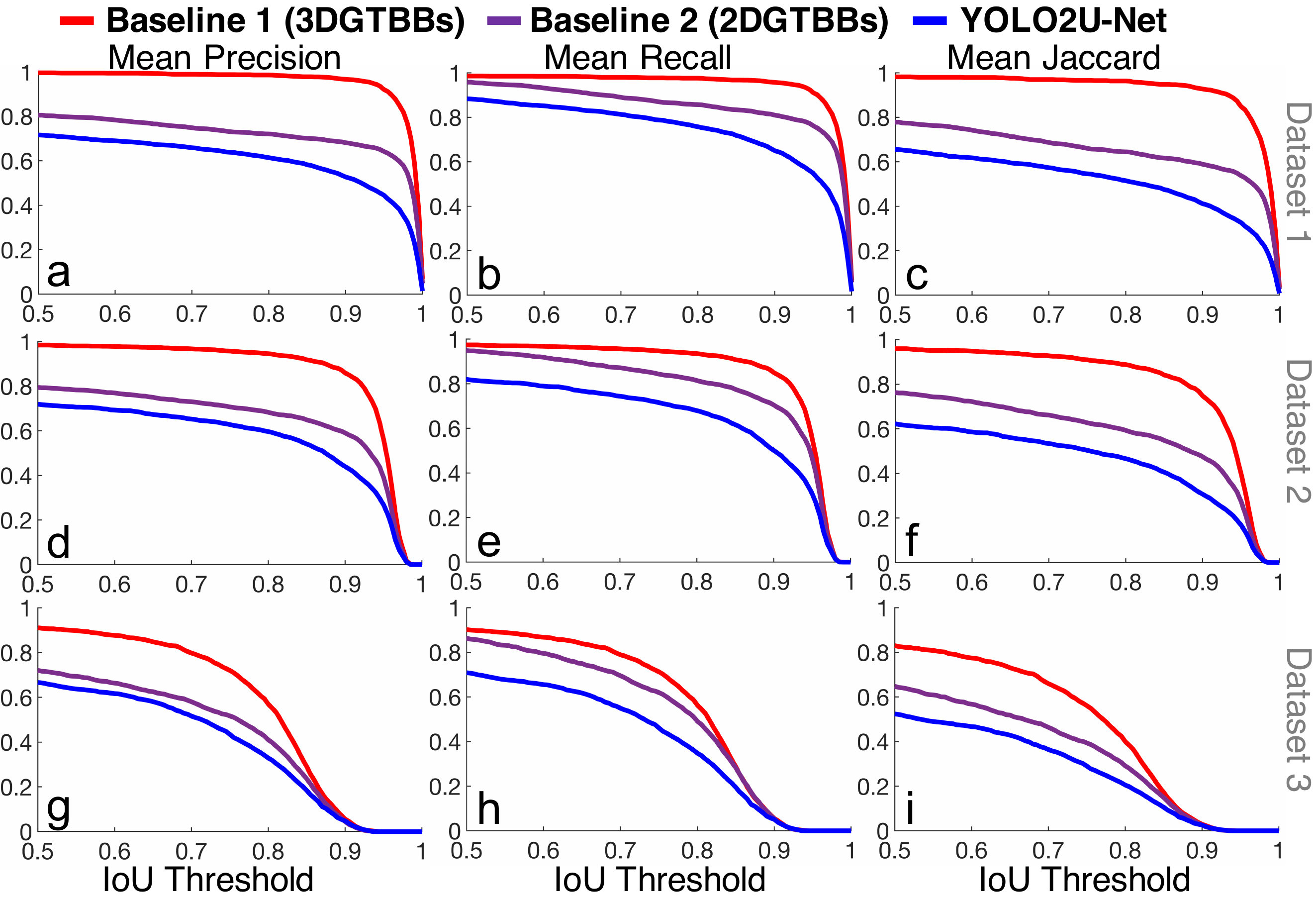}
\caption{Ablation study metrics to analyze the impact of YOLOv2 and the proposed fusion Algorithm~\ref{alg1} on the performance of YOLO2U-Net.
Baseline 1: using perfect 3D bounding boxes in last stage (3D U-Net).
Baseline 2: using perfect 2D bounding boxes (impact of box fusion).
Mean precision (a, d, g), recall (b, e, h), and Jaccard scores (c, f, i) for 20 volumes against IoU are plotted.}
\label{comp_perfect}
\end{figure}

%Three scenarios are considered. 
First, we consider the case that the 3D bounding boxes are perfectly known -- Baseline 1 (3DGTBBs). This is equivalent to a case in which YOLOv2 and Algorithm~\ref{alg1} both have perfect performance. 
Baseline 1 aims to show what the 3D U-Net can achieve using perfect inputs.
Second, we assume that only 2D bounding boxes of the cells are known -- Baseline 2 (2DGTBBs). Algorithm~\ref{alg1} is used to perform fusion of perfect 2D boxes and obtain 3D bounding boxes of the cells before inputting the data into 3D U-Net. 
Baseline 2 evaluates the impact of the fusion approach and algorithm. 
These baselines are used to directly compare with the full proposed YOLO2U-Net method as shown in Figure~\ref{comp_perfect}.

We compare the baselines to our full method for the three data sets and plot the average metric score as a function of IoU for each data set. 
We used three data sets with increasing complexity. 
In data set 1, we only added noise to simulated CompuCell3D data.; 
in 2, we added noise and Gaussian blur;
and in 3) we used realistic microscopy aberrations and noise (same data as in the previous sections).
It is evident from the figure that for the less challenging cases in data set 1 and 2 a near perfect score is obtainable by improving YOLOv2 (or using alternative methods) and the fusion algorithm. 
Further, enhancing just the fusion algorithm for these data sets will improve the cell counting accuracy (Jaccard score) for YOLO2U-Net.
For data set 3, even using perfect bounding boxes does not help in distinguishing and segmenting all cells correctly. 
This observation suggests that improvements to the current 3D U-Net are necessary for better instance segmentation of realistic data sets.
Table~\ref{Tab2} summarizes the mAJ values for the three scenarios discussed.
%GT stands for ground truth and BBs stands for bounding boxes.
This study clarifies that an improvement to YOLOv2 or replacement with a better 2D detection method can lead to significantly better performance with YOLO2U-Net.
The same argument is valid for the fusion algorithm.

\begin{table}[!htp]
\tiny
%\scalefont{0.5}
\centering
\caption{Ablation Study of {mAJ} Reliance on Architecture Components}
\label{Tab2}
%\resizebox{\columnwidth}{!}
\resizebox{.48\textwidth}{!}
{%
\begin{tabular}{|c|c|c|c|}
\hline
 & Baseline 1 (3DGTBBs) & Baseline 2 (2DGTBBs) & YOLO2U-Net\\
\hline
data set1 & 0.923   & 0.647  & 0.508    \\
\hline
data set2 & 0.798   & 0.567 & 0.442   \\
\hline
data set3 & 0.456   & 0.333  & 0.263   \\
\hline
\end{tabular}
}
\end{table}
\raggedbottom

\section{Conclusions and Future Work}\label{conclusions}

In this work, we've proposed a novel, versatile, and modular neural network architecture, which we called YOLO2U-Net, that combines two widely used deep learning architectures through an image processing-based fusion algorithm, and performs joint detection, localization, and 3D instance segmentation of cell nuclei.
The proposed method is a) efficient by localizing segmentation computation, b) adaptive to changes in object size through input re-scaling, and c) modular to enable plug-and-play future-proofing.
Several volumes of instance-level labeled data sets are simulated. 
These data sets challenge 3D instance segmentation models the same way real data does in two major aspects:  a) nontrivial cell geometry and cell-cell boundaries; and, b) out-of-plane signal mixing and low in-plane resolution. 
This data will be made publicly available.
To the best of our knowledge, such data sets containing characteristics of microscopy artifacts along with accurate instance segmentation masks are not publicly available.
The proposed method along with three 3D segmentation methods were trained and tested on the generated data sets.
In all cases, YOLO2U-Net outperform these current methods.
We also used an ablation study to analyse the impact of different components of the network on its performance. Given our reported findings, we are currently investigating integrating the components of the method into a model with full gradient path for end-to-end training to improve instance segmentation performance. 
To address lower performance on the most challenging data we are considering hyperparameter optimization techniques for the networks tested as well as drop-in improved networks as replacements. 
One such example of this is replacing the 3D U-Net with a Mixed Scale Dense Network proposed in~\cite{pelt2018mixed}.
For future work, we intend to expand the work for real data sets from different modalities and when needed use GANs and domain adaptation to improve the performance.
%\vspace{-0.4cm}
%\section{Compliance with Ethical Standards}
%This is a numerical simulation study for which no ethical approval was required.
\vspace{-0.4cm}
\section{Acknowledgments}
\label{sec:acknowledgments}

This collaborations was funded by St. Jude Children’s Research Hospital through funding from the American Lebanese Syrian Associated Charities (ALSAC). The Solecki Laboratory is funded by grants 1R01NS066936 and R01NS104029-02 from the National Institute of Neurological Disorders (NINDS).

\bibliographystyle{IEEEbib}
\bibliography{YOLO2UNet_main}

\end{document}